\documentclass[10pt,twocolumn]{revtex4}

\usepackage{graphicx}
\usepackage{epsfig}
\usepackage{hyperref}
\usepackage{dcolumn}
\usepackage{amssymb}
\usepackage{amsmath}
\usepackage{natbib}
\usepackage{braket}

\newcommand{\beqa}{\begin{eqnarray}}
\newcommand{\eeqa}{\end{eqnarray}}
\newcommand{\beq}{\begin{equation}}
\newcommand{\eeq}{\end{equation}}

\usepackage{color}

\begin{document}
\title{Excitation of an Atomic Transition with a Vortex Laser Beam}
\author{Christian T. Schmiegelow, Jonas Schulz, Henning Kaufmann, Thomas Ruster, Ulrich G. Poschinger and Ferdinand Schmidt-Kaler}
\affiliation{QUANTUM, Institut f\"ur Physik, Universit\"at Mainz, Staudingerweg 7, 55128 Mainz, Germany}

\maketitle

\textbf{Photons carry one unit of angular momentum associated with their spin~\cite{Beth1936}. Structured vortex beams carry additional orbital angular momentum which can also be transferred to matter~\cite{Allen1992}. This extra twist has been used for example to drive motion of microscopic particles in optical tweezers as well as to create vortices in degenerate quantum gases~\cite{He1995,Andersen2006}. Here we demonstrate the transfer of optical orbital angular momentum from the transverse spatial structure of the beam to the internal (electronic) degrees of freedom of an  atom. Probing a quadrupole transition of a single trapped $^{40}$Ca$^+$ ion  localized at the center of the vortex, we observe strongly modified selection rules, accounting for both the photon spin and the  vorticity of the field. In particular, we show that an atom can  absorb two quanta of angular momentum from a single photon even when rotational symmetry is conserved. In contrast to previous findings~\cite{Araoka2005,Loeffler2011a,Mathevet2013}, our experiment allows for conditions where the vorticity of the laser beam determines the optical excitation, contributing to the long-standing discussion on whether the orbital angular momentum of photons can be transferred to atomic internal degrees of  freedom~\cite{VanEnk1994,Babiker2002,Jauregui2004, Schmiegelow2012, Mondal2014, Scholz-Marggraf2014} and paves the way for its use to tailor light-matter interactions. 
Furthermore, we show that parasitic ac-Stark shifts from off-resonant transitions are suppressed in the center of the vortex beam, where the intensity vanishes but the field gradient is high. The mitigation of such shifts is of critical importance for the accuracy of optical clocks~\cite{Ludlow2015} and quantum logic gates~\cite{Haffner2003}. In the future, we expect these effects to be observed in molecules~\cite{Germann2014} as well as in artificial atoms~\cite{quinteiro2009electronic} which could lead to applications in many fields of research.}

\begin{figure}[hbp]\begin{center}
\includegraphics[width=0.45\textwidth]{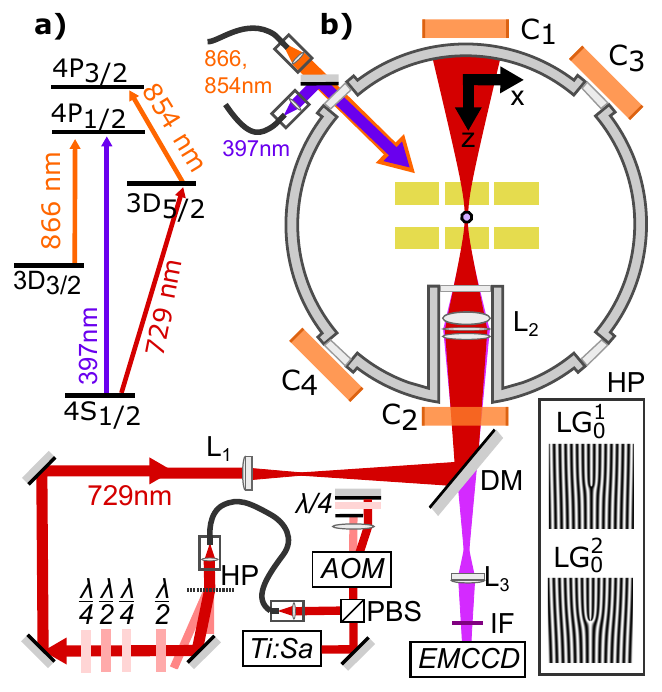}
\caption{a) Energy levels in $^{40}$Ca$^+$.  b) Experimental setup. Yellow bars indicate the linear segmented Paul trap inside an UHV chamber (gray) with an ion in the center. To excite the quadrupole $4^{2}S_{1/2}\leftrightarrow3^{2}D_{5/2}$ transition, coherent light from a Ti:Sa laser is transmitted through an acousto-optic modulator (AOM) for frequency and timing control, filtered by a polarization maintaining fibre and converted to the desired vortex beam with a holographic phase plate (HP). The laser beam polarization is set by a series of quarter- and half-wave plates, and focused onto the ion by lenses L$_1$ (f=50~mm) and L$_2$ (f=67~mm). Light resonant with the dipole transitions is used for Doppler cooling, detection (397 and 866~nm) and state reset (854~nm). Resonance fluorescence near 397~nm is imaged on an EMCCD camera with lenses L$_{2,3}$, passing a dichroic mirror (DM) and an interference filter (IF). The magnetic field is controlled by coils C$_{1-4}$ plus an additional coil (not shown) in the vertical direction.}
 \label{fig:sketch}
\end{center}\end{figure}

\begin{figure}[t]\begin{center}
\includegraphics[width=0.48\textwidth]{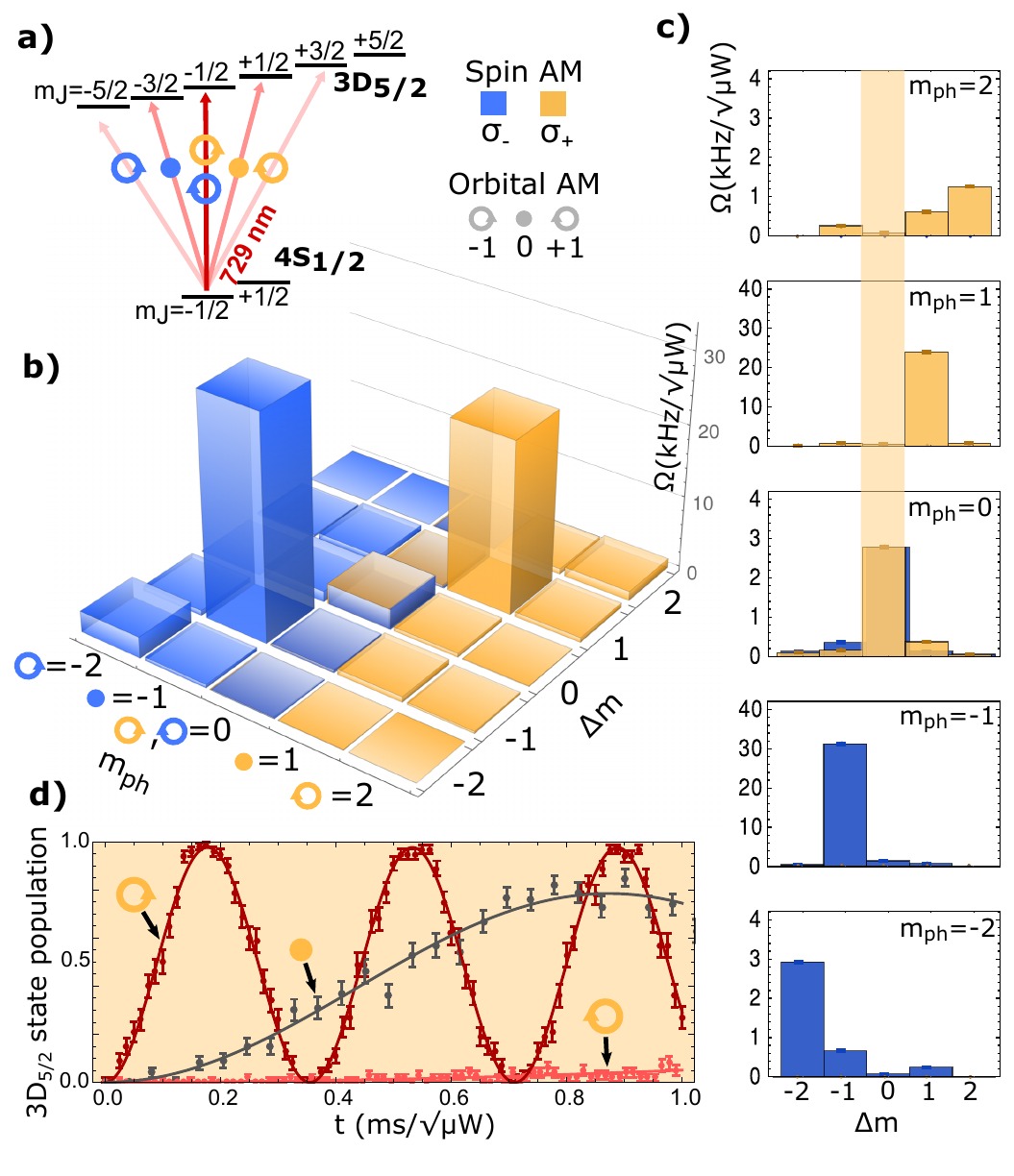}
\caption{a) Energy level structure for the 4$^{2}$S$_{1/2}\leftrightarrow\;$3$^{2}$D$_{5/2}$ manifold indicating which field structure and polarization drives each transition. Yellow (blue) indicates polarization $\sigma_{+(-)}$; clockwise (anti-clockwise) curl indicates vortex LG$_{0}^{+1}$(LG$_{0}^{-1}$) beams, dot indicates a Gaussian LG$_{0}^{0}$ beam. b,c) Interaction strength as Rabi frequencies for all spin and orbital angular momentum combinations and for all possible transitions, normalized to 1~$\mu$W of laser power. d) Example measurement data corresponding to $\sigma^{+}$, $\Delta m =0$ and all three beam combinations, indicated by the yellow stripe in c). For all experiments a total exposure time of 200~$\mu$s was scanned in steps of 5~$\mu$s for varying optical powers in the 3-60~$\mu$W range to determine the corresponding Rabi frequencies.}
\label{fig:rules}
\end{center}\end{figure}

The interaction between light and matter is governed by symmetries from which conservation laws of energy, momentum and angular momentum emerge. Electromagnetically driven transitions between two atomic states occur if the superposition of the charge distributions matches the multipole structure of the exciting field~\cite{Rochester2001}. This allows for categorizing atomic transitions in dipolar, quadrupolar and higher orders: the dipole moment couples to an oscillating electric field, while the quadrupole moment couples to an oscillating field gradient.  Here, we focus on quadrupole moments which can arise in electromagnetic fields under different conditions: in traveling plane waves, the oscillating longitudinal field gradient allows for driving electric quadrupole transitions. Alternatively, these transitions can also be driven by transverse gradients which occur e.g. in Laguerre-Gaussian LG$_{p}^{l}$ beams, which carry an additional orbital angular momentum (OAM) of $l$ per photon. Vortex beams with $\vert l\vert\geq 1$ exhibit a penumbra (dark center) where the intensity vanishes. Particularly strong transverse gradients at the penumbra occur in the case $l=\pm1$. LG beams have been proposed and employed for numerous novel applications~\cite{Torres2011, Andrews2013} such as high dimensional quantum information~\cite{Fickler2014}, quantum cryptography~\cite{Souza2008} and quantum memories~\cite{nicolas2014quantum}. Also, OAM-carrying beams of electrons~\cite{Verbeeck2010}, of neutrons~\cite{Clark2015} and of electromagnetic terahertz radiation~\cite{He2013} have been demonstrated.

For our experiments, we use a single laser-cooled $^{40}$Ca$^+$ ion trapped in a micro-structured segmented Paul trap\cite{Schulz2008} (see Fig.~\ref{fig:sketch}), which allows for precise positioning of the ion by changing the electrode voltages. A continuous-wave laser near 729~nm is used to drive the 4$^{2}$S$_{1/2}\leftrightarrow\;$3$^{2}$D$_{5/2}$ quadrupole transition. As indicated in Fig.~\ref{fig:sketch}, the laser is shaped to the transverse LG$_{0}^{0}$, LG$_{0}^{\pm1}$, and LG$_{0}^{\pm2}$ modes by holographic plates~\cite{Mair2001} and focused onto the ion with a beam waist of $w_0=$2.7(2)~$\mu$m. Upon exposure to the optical field on resonance with the quadrupole transition, the internal state of the ion undergoes coherent oscillations between the ground and excited state. The Rabi frequency $\Omega$ of these oscillations is measured to quantify the coupling strength. The Zeeman-split sub-levels of the 4$^{2}$S$_{1/2}\leftrightarrow\;$3$^{2}$D$_{5/2}$ transition are spectroscopically resolved due to an external magnetic field of 13~mT allowing to probe all transitions $\ket{4^{2}S_{1/2},m_J=\pm\tfrac{1}{2}}\leftrightarrow\ket{3^{2}D_{5/2},m_J=\pm\tfrac{1}{2},\pm\tfrac{3}{2},\pm\tfrac{5}{2}}$ independently by tuning the laser to the respective resonance (see methods).

We demonstrate the joint transfer of a quantum of orbital angular momentum and spin angular momentum (SAM) from the optical field to the ion. For this, we choose a setting with rotational symmetry about the propagation axis $z$ of the 729~nm beam. This is achieved by aligning the magnetic field along the $z$-direction and placing the ion in the beam center, at the bright center for the Gaussian LG$_{0}^{0}$ beam or at the dark penumbra for the LG$_{0}^{\pm1}$ beams, respectively. For this geometry, the angular momentum projection along $z$ is a conserved quantity, which enforces that transitions are allowed only if the total angular momentum $m_{ph}$  of the photon matches the difference in angular momentum projection $\Delta m$ between initial and final atomic states. The photon's total angular momentum $m_{ph}$ is given by the sum of SAM (for circular polarization $\sigma_{\pm}=\pm1$) and OAM ($l=0$ for LG$_{0}^{0}$ and $l=\pm1$ for LG$_{0}^{\pm1}$) modes. 

We verify that OAM contributes to angular momentum conservation by measuring the Rabi frequency for all possible values of $m_{ph}$ and $\Delta m$. We indeed observe coherent Rabi oscillations in all cases where angular momentum conservation is fulfilled: For $\Delta m=0$, an OAM of $\pm 1$ compensates for a SAM of $\mp 1$, while for $\Delta m=\pm 2$, OAM and SAM add up. In Fig.~\ref{fig:rules}, the results for initialization in the $\ket{4^{2}S_{1/2}, m_J=-\tfrac{1}{2}}$ state are shown, for the opposite spin initialization analogous results are obtained. For the transitions driven by the LG$_{0}^{0}$ (Gaussian) beam, the power-normalized Rabi frequencies are measured to be 13.0(8) times stronger than those driven by the LG$_{0}^{\pm1}$ (vortex) beams. This is consistent with the expected relative strength~\cite{Schmiegelow2012} for the measured beam waist, as determined by the ratio of waist to optical wavelength: $\pi w_0/\lambda=12.6(3)$. Additionally, the relative coupling strengths of different transitions are governed by the Wigner-Eckart theorem to account for coupling of SAM and OAM. For all transitions where angular momentum conservation is \textit{not} fulfilled, i.e. where $\Delta m\neq m_{ph}$, the measured coupling strengths are below 3\% of the coupling strength measured for the $m_{ph}=\Delta_m=\pm1$ transition, consistent with our error estimations (see methods).

For the case of an LG$_{0}^{\pm2}$ beam, we observe - within our experimental precision - negligible excitation for all transitions and polarization combinations. This is due to the fact that at the penumbra of the LG$_{0}^{2}$, the field amplitude increases quadratically in the radial direction. Thus, the quadrupole transition - which is driven by field gradients - cannot be excited at the center of an LG$_{0}^{\pm2}$ beam.

\begin{figure}[htp]\begin{center}
\includegraphics[width=0.48\textwidth]{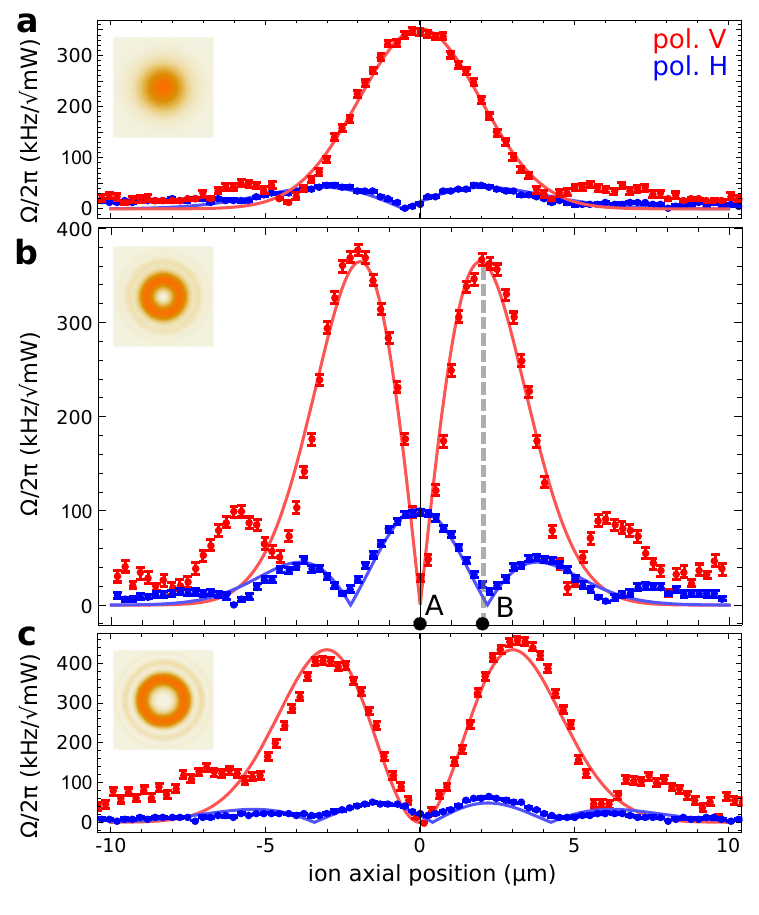}
\caption{Excitation profiles in units of power-normalized Rabi frequency as a function of the ion position across three different beams: a) LG$_{0}^{0}$ (Gaussian), b) LG$_{0}^{1}$ and c) LG$_{0}^{2}$.  Red (blue) data points correspond to $V$ ($H$) polarization, i.e. to excitation driven by the longitudinal (transverse) field gradient. Fits (solid lines) correspond to ideal LG beam intensity profiles (red) and their respective transverse field gradients (blue). The insets show beam intensity images taken with a CCD camera placed before the focusing lenses, revealing the same non-ideal LG outer ring structure as measured with the ion.
Note that in b), it can be clearly seen that the ion can be excited if it is in the dark penumbra of the LG$_{0}^{1}$ vortex beam. The measurements of the ac-Stark shift are carried out at positions marked $A$ and $B$. 
}
\label{fig:profiles}
\end{center}\end{figure}

Furthermore, we use the ion as a localized field probe~\cite{Guthohrlein2001,Horak2002} to map out the transverse and longitudinal field gradients for the LG$_{0}^{0}$, LG$_{0}^{1}$ and LG$_{0}^{2}$ beams by transversely moving the ion through the focus, see Fig.~\ref{fig:profiles}. Here, we choose the magnetic field at 45$^\circ$ with respect to the  propagation direction $z$ of the 729~nm beam, which is now linearly polarized. We probe the $\Delta m=\pm1$ transition, where the excitation mechanism is controlled by the  optical polarization~\cite{Schmiegelow2012} for this setting: if the electric field is  vertically ($V$) polarized (orthogonal to the plane spanned by magnetic field and beam propagation direction), the excitation strength is determined by the longitudinal gradient, proportional to the local intensity. By contrast, for horizontal ($H$) polarization, the coupling strength is given by the transverse gradient. 

For each of the LG beams, we observe that the Rabi frequency measurement with $V$ polarization reproduces the respective field amplitude profile, i.e. the coupling strength is proportional to the square root of the laser intensity. Conversely, for the case of the measurements with $H$ polarization, the strength of the transverse gradient is mapped out.  At the penumbra of the LG$_{0}^{1}$ vortex beam (Fig.~\ref{fig:profiles}b) the coupling mediated by the transverse gradient exceeds the one corresponding to the longitudinal gradient case by 6 standard deviations, i.e. the ion is actually \textit{excited in the dark}. In fact, this is the position at which the $\Delta m=0,\pm2$ transitions were driven in the previous experiment. By contrast, for the LG$_{0}^{\pm2}$ we observe that longitudinal- and transverse-gradient driven excitations are below our sensitivity limit at the center of the beam, see Fig.~\ref{fig:profiles}c. This shows, as mentioned before, that at the center of this beam both the electric field amplitude and its transverse gradient vanish. The outer shoulders observed in the measurements represent deviations from ideal Laguerre-Gaussian beam profiles. We confirm this by imaging the beam profiles on a CCD camera before the focusing lenses. The corresponding outer rings are clearly observed, see insets in Fig.\ref{fig:profiles}. These  beam imperfections result from fiber outcoupling and diffraction on the holographic plates generating the LG beams. 

An important challenge for the operation of laser-driven atomic qubits~\cite{Haffner2003} as well as for the implementation of optical frequency standards~\cite{Ludlow2015} is the mitigation of ac-Stark shifts. These are caused by off-resonant coupling of the probe beam to parasitic atomic transitions. In our setting, the ac-Stark shifts is mainly generated either by off-resonant driving of the quadrupole transition, or by far off-resonant coupling to the 4$^{2}$S$_{1/2}\leftrightarrow\;$4$^{2}$P$_{1/2}$, 4$^{2}$S$_{1/2}\leftrightarrow\;$4$^{2}$P$_{3/2}$ and 3$^{2}$D$_{5/2}\leftrightarrow\;$4$^{2}$P$_{3/2}$ dipole transitions. The quadrupolar ac Stark shift can be eliminated by polarization and magnetic field alignment, or by additional compensation beams. Here, we demonstrate that dipolar shift is substantially suppressed in the penumbra of a vortex beam, while the coherent coupling persists. 

We measure the energy shift $\Delta_S$ caused by the 729~nm beam in the LG$_{0}^{1}$ vortex mode, which is red detuned by $\Delta=2\pi\times$25~MHz with respect to the $\ket{4^{2}S_{1/2},m_J=+\tfrac{1}{2}}\leftrightarrow\ket{3^{2}D_{5/2},m_J=+\tfrac{3}{2}}$ transition. This is done by carrying out a Ramsey experiment, where a superposition of these two states acquires a phase $\phi=\Delta_S\times t$~\cite{Haffner2003} upon exposure to the beam at power $\mathcal{P}_{\Delta}$ for time $t$ (see methods). We compare the results for case $A$, where the ion is placed in the penumbra of the $H$ polarized beam, and for case $B$, where the ion placed at the intensity maximum of the side lobe of the $V$ polarized beam, as marked in Fig. \ref{fig:profiles}b). In case $A$, the transition is driven in the dark by the transverse gradient, whereas for case $B$, it is driven by the longitudinal gradient, and the ion is exposed to a high intensity. At position $A$, we measure an ac-Stark shift of $\Delta_S=2\pi\times$1.54(15)~kHz at a power of $\mathcal{P}_\Delta=$7.50(15)~mW and a Rabi frequency $\Omega=2\pi\times$11.93(24)~kHz at a power of $\mathcal{P}_\Delta=$20(1)~$\mu$W. At position $B$, we measure an ac-Stark shift of $\Delta_S=2\pi\times$19.1(1)~kHz at a power of $\mathcal{P}_\Delta=$1.75(4)~mW and a Rabi frequency $\Omega=2\pi\times$15.67(31)~kHz at a power of $\mathcal{P}_\Delta=$2.6(1)~$\mu$W. 

From the measured values for the ac-Stark shift, the power-normalized quadrupolar contribution is subtracted, i.e. $\Delta_S'=\Delta_S-(\mathcal{P}_{\Delta}/\mathcal{P}_{\Omega}) \Omega^2/(2\Delta)$. Additionally, the values are normalized by overall optical power $\mathcal{P}_{\Delta}$. The corrected, normalized shift in case $A$ is determined to be 0.70(25)\% of the corresponding value for case $B$. This residual shift is attributed to thermal fluctuations of the ion position into regions of nonzero field amplitude. Moreover, we compare the power-normalized ratios of the residual dipolar shift to the respective Rabi frequency. The quantity $\xi=(\Delta_S' \mathcal{P}_{\Delta}^{-1})/(\Omega\mathcal{P}_{\Omega}^{-1/2})$ is a suitable figure-of-merit, as it gives the ratio of the undesired shift to the desired coherent coupling when multiplied with the square root of the employed optical power. For the transverse-gradient driven excitation in the dark, we find a suppression by $\xi_A/\xi_B=$2.5(9)\% as compared to the longitudinal-gradient driven case. 

Our results open up a wide range of future research directions, in which OAM of light can be harnessed as new control parameter in light-matter interaction. Possible applications include the mapping of high dimensional flying qubits from photons to atomic quantum memories~\cite{Fickler2014}, tailored  interactions to improve quantum logic gates~\cite{Haffner2003}, the control of quantized motion of trapped particles~\cite{Mondal2014, Schmiegelow2012} and the suppression of ac-Stark shifts for optical clock transitions~\cite{Ludlow2015}.  The key aspects in our study are the use of a quadrupole transition, focusing the probe beam close to the diffraction limit and using a well-localized atomic system. Determining which of these conditions are sufficient to observe enhanced effects due to the structure of the beam is a prerequisite to extending this technique to other physical systems such as molecules or artificial atoms.

\textbf{Acknowledgements}\\
We thank Rupert Ursin for lending us the holographic phase plates; A. Wiens and A. Walther for contributions on early stages of the experiment, S. Franke-Arnold and D. Budker for useful comments, and A. Z. Khoury for inspiring this work with a lecture on orbital angular momentum of photons at the J. A. Swieca School in 2008 in Sao Paulo. C.T.S. acknowledges the support of the Alexander von Humboldt Foundation. 

\textbf{Author Contributions}  
C.T.S. and F.S.K. conceived the idea of the experiment. Experimental data was
taken by C.T.S. and J.S, using an apparatus primarily set up by C.T.S., J.S., H.K. and T.R.. 
Data analysis was performed by C.T.S, J.S. and U.G.P. The paper was written by C.T.S.,
F.S.K., J.S. and U.G.P., with input from all authors.

\bibliographystyle{unsrt}
\bibliography{references}  

\begin{thebibliography}{10}

\bibitem{Beth1936}
R.~A. Beth.
\newblock {Mechanical detection and measurement of the angular momentum of
  light}.
\newblock {\em Physical Review}, 50(2):115--125, 1936.

\bibitem{Allen1992}
L.~Allen, M.W. Beijersbergen, R.J.C. Spreeuw, and J.P. Woerdman.
\newblock {Orbital angular momentum of light and the transformation of
  Laguerre-Gaussian laser modes}.
\newblock {\em Physical Review A}, 45(11):8185--8189, 1992.

\bibitem{He1995}
H.~He, M.E.J. Friese, N.R. Heckenberg, and H.~Rubinsztein-Dunlop.
\newblock {Direct Observation of Transfer of Angular Momentum to Absorptive
  Particles from a Laser Beam with a Phase Singularity}.
\newblock {\em Physical Review Letters}, 75(5):826--829, 1995.

\bibitem{Andersen2006}
M.~F. Andersen, C.~Ryu, Pierre Clad\'{e}, Vasant Natarajan, a.~Vaziri,
  K.~Helmerson, and W.~D. Phillips.
\newblock {Quantized rotation of atoms from photons with orbital angular
  momentum}.
\newblock {\em Physical Review Letters}, 97(17):170406, 2006.

\bibitem{Araoka2005}
F.~Araoka, T.~Verbiest, K.~Clays, and A.~Persoons.
\newblock {Interactions of twisted light with chiral molecules: An experimental
  investigation}.
\newblock {\em Physical Review A - Atomic, Molecular, and Optical Physics},
  71(5):1--3, 2005.

\bibitem{Loeffler2011a}
W.~L\"{o}ffler, D.J. Broer, and J.P. Woerdman.
\newblock {Circular dichroism of cholesteric polymers and the orbital angular
  momentum of light}.
\newblock {\em Physical Review A - Atomic, Molecular, and Optical Physics},
  83(6):1--3, 2011.

\bibitem{Mathevet2013}
R.~Mathevet, B.V. de~Lesegno, L.~Pruvost, and G.L.J.A. Rikken.
\newblock {Negative experimental evidence for magneto-orbital dichroism.}
\newblock {\em Optics express}, 21(4):3941--5, 2013.

\bibitem{VanEnk1994}
S.J. {Van Enk} and G.~Nienhuis.
\newblock {Commutation Rules and Eigenvalues of Spin and Orbital Angular
  Momentum of Radiation Fields}.
\newblock {\em Journal of Modern Optics}, 41(5):963--977, 1994.

\bibitem{Babiker2002}
M.~Babiker, C.R. Bennett, D.L. Andrews, and L.C. {D\'{a}vila Romero}.
\newblock {Orbital angular momentum exchange in the interaction of twisted
  light with molecules.}
\newblock {\em Physical Review Letters}, 89(14):143601, 2002.

\bibitem{Jauregui2004}
{R. Jauregui}.
\newblock {Rotational effects of twisted light on atoms beyond the paraxial
  approximation}.
\newblock {\em Physical Review A}, 033415(70):033415, 2004.

\bibitem{Schmiegelow2012}
C.T. Schmiegelow and F.~Schmidt-Kaler.
\newblock {Light with orbital angular momentum interacting with trapped ions}.
\newblock {\em European Physical Journal D}, 66(6):1--9, 2012.

\bibitem{Mondal2014}
P.K. Mondal, B.~Deb, and S.~Majumder.
\newblock {Angular momentum transfer in interaction of Laguerre-Gaussian beams
  with atoms and molecules}.
\newblock {\em Physical Review A - Atomic, Molecular, and Optical Physics},
  89:29--33, 2014.

\bibitem{Scholz-Marggraf2014}
H.~M. Scholz-Marggraf, S.~Fritzsche, V.~G. Serbo, A.~Afanasev, and
  A.~Surzhykov.
\newblock {Absorption of twisted light by hydrogenlike atoms}.
\newblock {\em Physical Review A - Atomic, Molecular, and Optical Physics},
  90:013425, 2014.

\bibitem{Ludlow2015}
A.D. Ludlow, M.M. Boyd, J.~Ye, E.~Peik, and P.O. Schmidt.
\newblock {Optical atomic clocks}.
\newblock {\em Reviews of Modern Physics}, 87(2):637--701, 2015.

\bibitem{Haffner2003}
H.~H\"{a}ffner, S.~Gulde, M.~Riebe, G.~Lancaster, C.~Becher, J.~Eschner,
  F.~Schmidt-Kaler, and R.~Blatt.
\newblock {Precision measurement and compensation of optical stark shifts for
  an ion-trap quantum processor.}
\newblock {\em Physical Review Letters}, 90:143602, 2003.

\bibitem{Germann2014}
M.~Germann, X.~Tong, and S.~Willitsch.
\newblock {Observation of electric-dipole-forbidden infrared transitions in
  cold molecular ions}.
\newblock {\em Nature Physics}, 10(11):820--824, 2014.

\bibitem{quinteiro2009electronic}
G.F. Quinteiro and P.I. Tamborenea.
\newblock {Electronic transitions in disk-shaped quantum dots induced by
  twisted light}.
\newblock {\em Physical Review B}, 79(15):155450, 2009.

\bibitem{Rochester2001}
S.M. Rochester and D.~Budker.
\newblock {Atomic polarization visualized}.
\newblock {\em American Journal of Physics}, 69(4):450, 2001.

\bibitem{Torres2011}
J.P. Torres and L.~Torner.
\newblock {\em Twisted Photons: Applications of Light with Orbital Angular
  Momentum}.
\newblock Wiley, 2011.

\bibitem{Andrews2013}
D.L. Andrews and M.~Babiker.
\newblock {\em The Angular Momentum of Light}.
\newblock Cambridge University Press, 2013.

\bibitem{Fickler2014}
R.~Fickler, R.~Lapkiewicz, M.~Huber, M.P.J. Lavery, M.J. Padgett, and
  A.~Zeilinger.
\newblock {Interface between path and orbital angular momentum entanglement for
  high-dimensional photonic quantum information}.
\newblock {\em Nature Communications}, 5:5, 2014.

\bibitem{Souza2008}
C.~E.~R. Souza, C.~V.~S. Borges, A.~Z. Khoury, J.~A.~O. Huguenin, L.~Aolita,
  and S.~P. Walborn.
\newblock Quantum key distribution without a shared reference frame.
\newblock {\em Phys. Rev. A}, 77:032345, Mar 2008.

\bibitem{nicolas2014quantum}
A.~Nicolas, L.~Veissier, L.~Giner, E.~Giacobino, D.~Maxein, and J.~Laurat.
\newblock A quantum memory for orbital angular momentum photonic qubits.
\newblock {\em Nature Photonics}, 8(3):234--238, 2014.

\bibitem{Verbeeck2010}
J.~Verbeeck, H.~Tian, and P.~Schattschneider.
\newblock {Production and application of electron vortex beams}.
\newblock {\em Nature}, 467(7313):301--304, 2010.

\bibitem{Clark2015}
C.W. Clark, R.~Barankov, M.G. Huber, M.~Arif, D.G. Cory, and D.A. Pushin.
\newblock {Controlling neutron orbital angular momentum}.
\newblock {\em Nature}, 525(7570):504--506, 2015.

\bibitem{He2013}
J.~He, X.~Wang, D.~Hu, J.~Ye, S.~Feng, Q.~Kan, and Y.~Zhang.
\newblock {Generation and evolution of the terahertz vortex beam}.
\newblock {\em Optics Express}, 21(17):20230, 2013.

\bibitem{Schulz2008}
S.A. Schulz, U.~Poschinger, F.~Ziesel, and F.~Schmidt-Kaler.
\newblock {Sideband cooling and coherent dynamics in a microchip
  multi-segmented ion trap}.
\newblock {\em New Journal of Physics}, 10, 2008.

\bibitem{Mair2001}
A.~Mair, A.~Vaziri, G.~Weihs, and A.~Zeilinger.
\newblock {Entanglement of the orbital angular momentum states of photons.}
\newblock {\em Nature}, 412(6844):313--316, 2001.

\bibitem{Guthohrlein2001}
G.R. Guth\"{o}hrlein, M.~Keller, K.~Hayasaka, W.~Lange, and H.~Walther.
\newblock {A single ion as a nanoscopic probe of an optical field.}
\newblock {\em Nature}, 414(6859):49--51, 2001.

\bibitem{Horak2002}
P.~Horak, H.~Ritsch, T.~Fischer, P.~Maunz, T.~Puppe, P.W.H. Pinkse, and
  G.~Rempe.
\newblock {Optical Kaleidoscope Using a Single Atom}.
\newblock {\em Physical Review Letters}, 88(4):043601, 2002.

\end{thebibliography}

\begin{center}
\textbf{Methods}
\end{center}
\textbf{Coupling strength determination sequence.} Each sequence starts with Doppler laser cooling, followed by optical pumping into either of the two ground state sublevels $\ket{4^{2}S_{1/2}, m_S=\pm\tfrac{1}{2}}$.  Next, the probe pulse near 729~nm is applied for driving Rabi oscillations between the ground state and the metastable 3$^{2}$D$_{5/2}$ state. The final state is determined from observing state-dependent fluorescence on an EMCCD camera while illuminating the ion near 397~nm and 866~nm. Observation of fluorescence indicates the ion to be in the $S_{1/2}$ state, while absence of fluorescence indicates a collapse into the 3$^{2}$D$_{5/2}$ state. Before the sequence is repeated, we apply light near 854~nm to remove population from the 3$^{2}$D$_{5/2}$ manifold, see Fig. \ref{fig:sketch}. By repeating this sequence 200 times, we obtain an estimate of the 3$^{2}$D$_{5/2}$ state occupation probability. By measuring this excitation probability versus the probe pulse duration, ranging up to a few hundred $\mu$s, we determine the Rabi frequency $\Omega$. 

\textbf{Error estimation on the coupling strengths.} Small residual excitation measured on forbidden transitions, where $\Delta m\neq m_{ph}$, is attributed to three effects: thermal position fluctuations of the ion, imperfect optical polarization and a non-zero angle between the magnetic field and the laser propagation direction.  The thermal position spread is most prominent on the transitions involving the vortex beam due to its sharp transverse structure ($m_{ph}=0,\pm2$, see Fig.\ref{fig:rules}c). To estimate the excess excitation, we calculate the overlap between the beam's field profile with the ion´s thermal position spread. This spread is given by about 60~nm, as independently measured for our experimental conditions of a Doppler-cooled $^{40}$Ca$^+$ ion. Consistent with these estimations, all spurious couplings observed are below 3\% of the coupling strength pertaining to the $m_{ph}=\Delta m=\pm1$ transition. 

\textbf{Stark shift determination sequence.} State preparation and readout are carried out as for the sequence before. After preparation, the ion located either at the penumbra or at the intensity maximum is exposed to the probe beam resonant to the $\ket{4^{2}S_{1/2},m_J=+\tfrac{1}{2}}\leftrightarrow\ket{3^{2}D_{5/2},m_J=+\tfrac{3}{2}}$ at a pulse area of $\pi/2$, such that a balanced superposition of both states is created. Next, it is exposed to the off-resonant vortex beam for time $t$, which induces an ac-Stark phase shift on the superposition. Finally, a second resonant $\pi/2$ pulse is applied. After recording the final population in the excited state versus $t$, the ac-Stark shift is determined by the frequency of the resulting coherent oscillations.

\textbf{Optical pumping.} For the beam profile reconstructions and the ac-Stark shift experiments, where the magnetic field was at 45$^\circ$ with respect to the 729~nm beam, optical pumping was carried out with $\sigma_+$ polarized light driving the 4$^{2}$S$_{1/2}\leftrightarrow\;$4$^{2}$P$_{1/2}$ dipole transition near 397~nm. For the experiments on the determination of the transition selection rules, the 729~nm beam is aligned parallel to the magnetic field. Here, pumping is carried out by transferring population from the 4$^{2}$S$_{1/2}$ levels to be depleted to a 3$^{2}$D$_{5/2}$ level and then resetting the population to the 4$^{2}$S$_{1/2}$ manifold with light resonant on the 3$^{2}$D$_{5/2}\leftrightarrow\;$4$^{2}$P$_{3/2}$ transition near 854~nm, see Fig. \ref{fig:sketch}. By repeating this sequence ~10 times, we can prepare the desired $\ket{4^{2}S_{1/2},m_J=\pm\tfrac{1}{2}}$ state at a fidelity $\gtrsim$99\%.

\end{document}